\documentclass[fleqn,usenatbib]{mnras}

\usepackage{newtxtext,newtxmath}

\usepackage[T1]{fontenc}

\DeclareRobustCommand{\VAN}[3]{#2}
\let\VANthebibliography\thebibliography
\def\thebibliography{\DeclareRobustCommand{\VAN}[3]{##3}\VANthebibliography}

\usepackage{graphicx}	
\usepackage{amsmath}	

\usepackage{amssymb}	

\title[\textit{Kepler-TESS} ephemerides update]{Revisiting the \textit{Kepler} field with \textit{TESS}: Improved ephemerides using \textit{TESS} 2min data}

\author[M. P. Battley et al.]{
Matthew P. Battley,$^{1,2}$\thanks{E-mail: Matthew.Battley@warwick.ac.uk}
Michelle Kunimoto,$^{3}$
David J. Armstrong,$^{1,2}$
Don Pollacco,$^{1,2}$
\\
$^{1}$Dept. of Physics, University of Warwick, Gibbet Hill Road, Coventry CV4 7AL, UK\\
$^{2}$Centre for Exoplanets and Habitability, University of Warwick, Gibbet Hill Road, Coventry CV4 7AL, UK\\
$^{3}$ Kavli Institute for Astrophysics and Space Research, Massachusetts Institute of Technology, Cambridge, MA 02139 
}

\date{Accepted XXX. Received YYY; in original form ZZZ}

\pubyear{2021}

\begin{document}
\label{firstpage}
\pagerange{\pageref{firstpage}--\pageref{lastpage}}
\maketitle

\begin{abstract}
Up to date planet ephemerides are becoming increasingly important as exoplanet science moves from detecting exoplanets to characterising their architectures and atmospheres in depth. In this work ephemerides are updated for 22 \textit{Kepler} planets and 4 \textit{Kepler} planet candidates, constituting all \textit{Kepler} planets and candidates with sufficient signal to noise in the \textit{TESS} 2min dataset. A purely photometric method is utilised here to allow ephemeris updates for planets even when they do not posses significant radial velocity data. The obtained ephemerides are of very high precision and at least seven years 'fresher' than archival ephemerides. In particular, significantly reduced period uncertainties for Kepler-411d, Kepler-538b and the candidates K00075.01/K00076.01 are reported. O-C diagrams were generated for all objects, with the most interesting ones discussed here. Updated TTV fits of five known multiplanet systems with significant TTVs were also attempted (Kepler-18, Kepler-25, Kepler-51, Kepler-89, and Kepler-396), however these suffered from the comparative scarcity and dimness of these systems in \textit{TESS}. Despite these difficulties, \textit{TESS} has once again shown itself to be an incredibly powerful follow-up instrument as well as a planet-finder in its own right. Extension of the methods used in this paper to the 30min-cadence \textit{TESS} data and \textit{TESS} extended mission has the potential to yield updated ephemerides of hundreds more systems in the future.

\end{abstract}




\section{Introduction}

Maintenance of planet ephemerides is crucial to further characterisation of known planets. This is especially true for follow-up with high-profile observatories such as the upcoming \textit{James Webb Space Telescope} \citep[\textit{JWST},][]{Gardner2006TheTelescope}, where timing uncertainties of less than 30 minutes are desirable \citep{Dragomir2020SecuringEphemerides}. Furthermore, ensuring that ephemerides of known planets and planet candidates remain fresh secures the legacy of large-scale planet-finding missions such as those performed by the \textit{Kepler} satellite \citep{Borucki2010KeplerResults} and the \textit{Transiting Exoplanet Survey Satellite} \citep[\textit{TESS},][]{Rickeretal.2014TheSatellite}. 

The \textit{Kepler} satellite dramatically changed the field of exoplanet science, discovering almost 3000 validated exoplanets and thousands more planet candidates over the course of its main mission and following \textit{K2} mission \citep{Howell2014TheResults}. However, the original \textit{Kepler} mission finished May 11 2013 after the failure of two of the satellite's reaction wheels\footnote{https://archive.stsci.edu/missions/kepler/docs/drn/release\_notes25/KSCI-19065-002DRN25.pdf}, meaning that it has now been over seven years since most of these planets have been observed. This observational gap, coupled with uncertainties in the periods and epochs of the \textit{Kepler} transits, has led to many of the planet/candidate ephemerides becoming imprecise or 'stale'.

A number of different surveys have attempted to solve this problem, both for \textit{Kepler} and \textit{K2} planets \citep[e.g.][]{Livingston2019SpitzerMission,Ikwut-Ukwa2020TheK2-261,Edwards2020OriginalExoplanets}, however the sheer number of planets and candidates discovered by the satellite makes this a significant challenge. Promising follow-up solutions include re-observing the entire \textit{Kepler} field or using large-scale citizen-science approaches \citep[e.g.][]{Zellem2020UtilizingFollow-up,Kokori2020ExoClockPublic}. The launch of \textit{TESS} in 2018 provides a new opportunity to address this challenge. Having already completed its two-year primary mission, \textit{TESS} has surveyed \textasciitilde75\% of the night sky. During this mission, \textit{TESS} returned to the original \textit{Kepler} field, providing new data for the \textit{Kepler} field in Sector 14 (and to a lesser extent, Sectors 15 and 26) of its observations. 

\citet{Ikwut-Ukwa2020TheK2-261} have previously demonstrated the significant impact re-observing data from the \textit{Kepler} satellite with \textit{TESS} can have, completing full ephemeris and wider system parameter updates for K2-114, K2-167, K2-237 and K2-261 by utilising a combined photometric (\textit{TESS} and \textit{Kepler} data) and spectroscopic (using archival radial velocities) measurements. In the case of K2-114 these new measurements were shown to reduce the uncertainty in the planet's period by a factor of 66 \citep{Ikwut-Ukwa2020TheK2-261} compared to its discovery period. This promising result heralds the way for a wider treatment of the entire \textit{TESS/Kepler} crossover sample in order to have ephemerides ready for the launch of \textit{JWST}.  

Furthermore, \citet{Christ2018ObservationsFeatures} point out that re-observing the \textit{Kepler} field with \textit{TESS} provides a host of additional benefits, providing a window into long-term trends such as tidal decay of Hot Jupiters. In particular, \citet{Christ2018ObservationsFeatures} suggest that the Hot Jupiters Kepler-2b/HAT-P-7 and KOI-13b will be particularly interesting objects to follow up with \textit{TESS} data. They also suggest that \textit{TESS} may be very powerful for investigating longer-term transit timing variations. These thoughts are echoed by the modelling of \citet{Goldberg2018ProspectsObservations} where they conclude that mass uncertainties could be improved for 6-14 planets with the new \textit{TESS} 2min data, depending on the eventual measurement uncertainties.  

While the combined photometric/spectroscopic approach of \citet{Ikwut-Ukwa2020TheK2-261} allowed for new global models of the systems to be created, many of the lower-mass Kepler planets and candidates do not currently possess significant radial-velocity data. This motivates the construction of a homogeneous method of updating new ephemerides based on \textit{TESS/Kepler} photometry alone. Such a method is presented in this paper, along with the updated ephemerides for every \textit{Kepler} planet and candidate which was reasonably re-observed in the \textit{TESS} 2min data.

Section \ref{observations} describes the \textit{Kepler} and \textit{TESS} observations used in this analysis. This is followed by the methods used to obtain the individual transit times and updated ephemerides in Section \ref{methods} before the results are presented in Section \ref{results}. Implications of these results are discussed further in Section \ref{discussion} before a summary of the work carried out. Note that a wider list of all \textit{Kepler} planetary systems which received \textit{TESS} short-cadence data is included in the appendix, including reasons why excluded systems were removed from the analysis. 

\section{Observations} \label{observations}

\subsection{\textit{Kepler}} \label{Kepler_phot}

\textit{Kepler}'s primary mission ran for approximately four years, or 17 'quarters', from first light on May 2nd 2009 until the loss of the satellite's second reaction wheel in May 2013. Observing the same rich patch of sky in the vicinity of the Cygnus and Lyra constellations for its entire primary mission, the \textit{Kepler} satellite yielded an unprecedented volume of high quality, long-duration photometry, and was the first telescope capable of finding Earth-size planets around Sun-like stars in year-long orbits \citep{Borucki2010KeplerResults}. The majority of the targets in the \textit{Kepler} mission's primary field were observed in \textit{Kepler's} 29.4min (30min) long-cadence mode, however approximately 512 objects per quarter received 1min short-cadence light-curves covering a month in time \citep{Thompson2016KeplerManual}. 

In this work, \textit{Kepler} long-cadence PDCSAP light-curves were retrieved from the public Mikulski Archive for Space Telescopes (MAST)\footnote{https://mast.stsci.edu/portal/Mashup/Clients/Mast/Portal.html} using the Lightkurve python package \citep{LightkurveCollaboration2018Lightkurve:Python}. These light-curves were prepared by the standard \textit{Kepler} science processing pipeline \citep{Jenkins2010OverviewPipeline}, which involves pixel-level calibration, smear and background removal, optimal aperture selection and modelling of systematic errors introduced by the spacecraft. Due to the stellar variability of most sources over the long \textit{Kepler} observation timeline, an additional detrending step was necessary before planetary transits could be used to update ephemerides. This was achieved by applying a simple 24hr window LOWESS filter \citep{Cleveland1979RobustScatterplots,Battley2020AImages} to the out-of-transit light-curve, except in the cases such as Kepler-9 \citep{Holman2010Kepler-9:Variations}, where shorter 12-15hr windows were required to handle the shorter-period activity cycles. In order to preserve the form of each individual transit, the transit epochs were masked from the light-curves during this detrending step and replaced with linear interpolations until the detrending was complete. 

\subsection{\textit{TESS}} \label{TESS_phot}

\begin{figure}
    \includegraphics[width=\columnwidth]{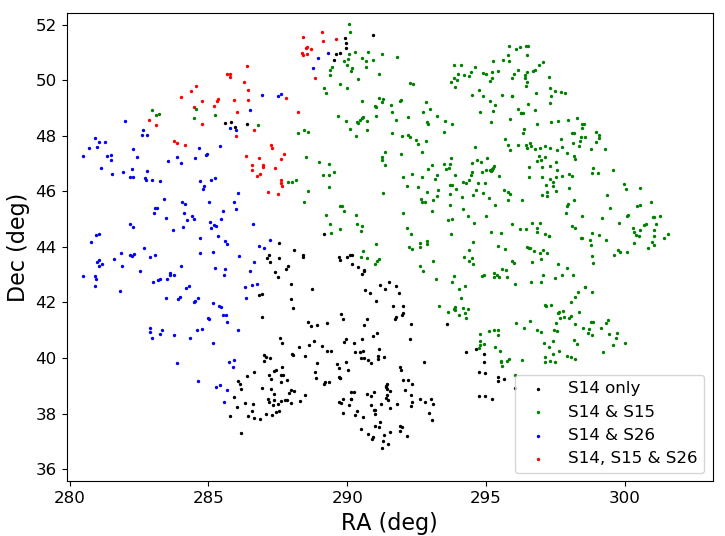}
    \caption{Overview of \textit{Kepler} stars reobserved by \textit{TESS} with 2min cadence. Individual stars are colour-coded according to which \textit{TESS} sectors they were observed in: Black: Sector 14 only; Green: Sectors 14 and 15; Blue: Sectors 14 and 26; Red: Sectors 14, 15 and 26.}
     \label{fig:observations_overview}
 \end{figure}

\textit{TESS} re-observed the \textit{Kepler} field in the second year of its primary mission. The majority of the Kepler field was observed in Sector 14 (between 18 July to 15 August 2019), but due to overlap in adjacent \textit{TESS} observations, some of these stars were also re-observed in Sectors 15 (15 August - 11th September 2019) and 26 (8 June - 4th July 2020). However, although the entire \textit{Kepler} field was re-observed in Sector 14 of the \text{TESS} primary mission, only 962 Kepler stars were pre-selected to receive 2min data in \textit{TESS}'s primary mission. These stars are plotted in Figure \ref{fig:observations_overview}, colour-coded by which sectors they were re-observed in. 

All \textit{TESS} light-curves used in this work were generated by NASA's Science Processing Operations Centre (SPOC) and, similarly to the \textit{Kepler} data, accessed using the Lightkurve python package \citep{LightkurveCollaboration2018Lightkurve:Python}. This data was extracted from the raw images using the standard SPOC pipeline \citep{Jenkins2016TheCenter}, which includes pixel-level decorrelation, centroiding and aperture optimisation. For all light-curves the  PDCSAP flux data was analysed, which had previously had systematic errors such as times of poor pointing or excessive scattered light removed \citep{Jenkins2016TheCenter}. Similar to the \textit{Kepler} data, after extraction via the Lightkurve package, long-term stellar variability outside the transit was removed from the \textit{TESS} 2min light-curves using a 24hr window LOWESS filter. 

\subsection{\textit{Photometry Comparison}} \label{comparison}

\begin{figure*}
    \includegraphics[width=\textwidth]{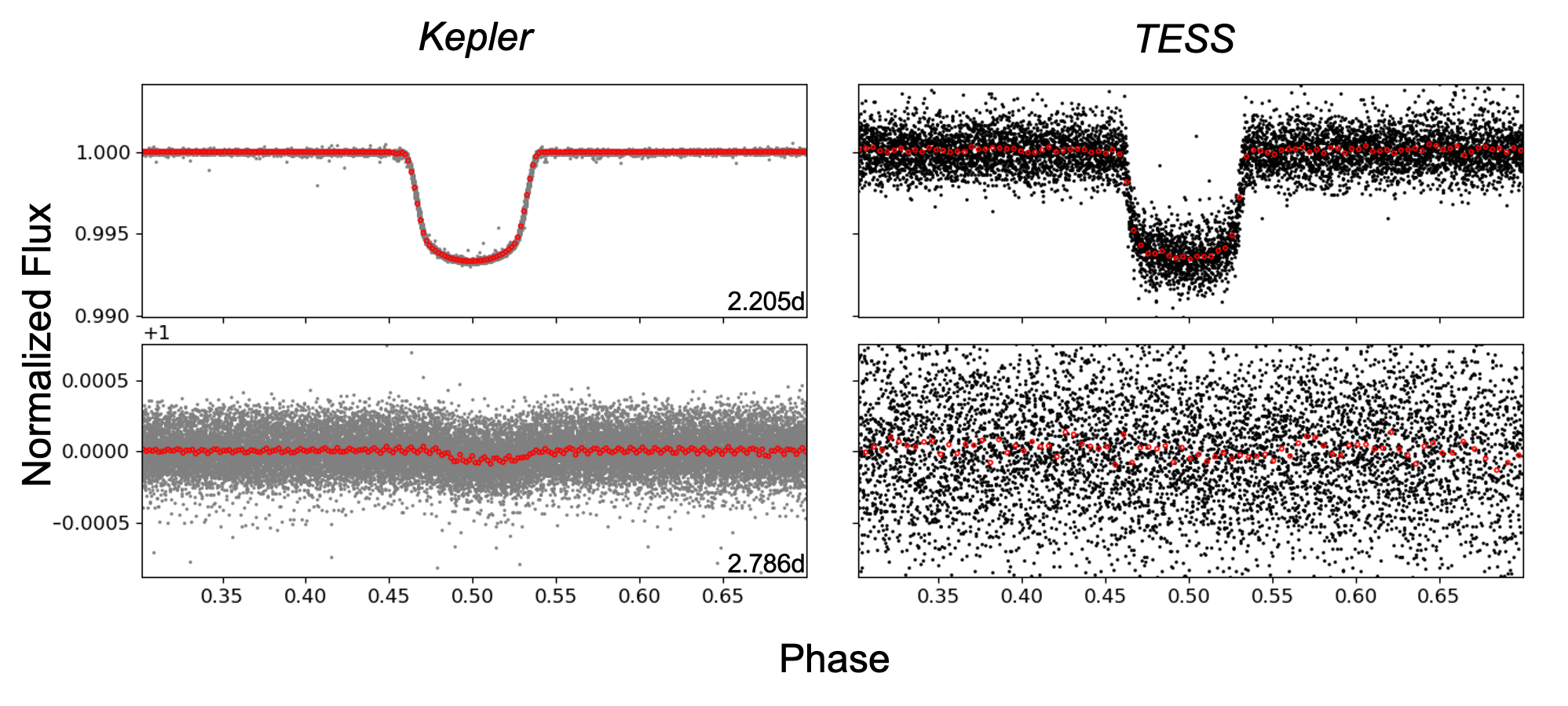}
    \caption{Photometric data comparison between \textit{Kepler} (left) and \textit{TESS} (right). Top: HAT-P-7 b/\textit{Kepler}-2 b, a 1.51 $M_{Jup}$ Hot Jupiter is clearly recovered by both \textit{Kepler} and \textit{TESS}. Bottom: \textit{Kepler-21} b, a 1.6$M_{\bigoplus}$ super-Earth is clear in the folded \textit{Kepler} data but lost in scatter in the \textit{TESS} data.}
     \label{fig:phot_comparison}
 \end{figure*}

The \textit{Kepler} and \textit{TESS} missions were designed with two very different survey strategies in mind, and hence differ in their photometric performance. While the \textit{Kepler} mission focussed on a relatively small (100 deg$^2$, or ~0.25\% of the sky) northern hemisphere section of the sky near the Cygnus, Lyra and Draco constellations \citep{Borucki2010KeplerResults}, the \textit{TESS} primary mission focussed on maximising the amount of sky viewed, achieving almost 75\% coverage of the sky\footnote{https://www.nasa.gov/feature/goddard2020/nasa-s-planet-hunter-completes-its-primary-mission} in its two year primary mission, updated from the originally planned 85\% due to issues with scattered light \citep{Rickeretal.2014TheSatellite}. To achieve this coverage with \textit{TESS}, a slight sacrifice in sensitivity was required, with combined differential photometric precision (CDPP) dropping from an average of 30ppm for the \textit{Kepler} mission \citet{Gilliland2011KeplerProperties} to approximately 100ppm\footnote{https://archive.stsci.edu/missions/tess/doc/tess\_drn/tess\_sector\\ \_14\_drn19\_v02.pdf} for Sector 14 of the \textit{TESS} mission. This reduced sensitivity comes from a variety of sources, but most importantly the reduced effective lens size in \textit{TESS} due to reducing from a single large mirror (effictive lens diameter of 95cm) on the \textit{Kepler} satellite to smaller 10.5cm lens-based telsecopes in \textit{TESS}. This reduction in effective lens diameter results in the lower resolution of the \textit{TESS} instruments (21 arcseconds per pixel instead of Kepler's 4 arcseconds per pixel) and yields increased blending from nearby stars compared to the \textit{Kepler} satellite. The combination of these effects, coupled with the shorter observational timescale in \textit{TESS} (27 days to 1 year in \textit{TESS}; 4 years in \textit{Kepler}), means that \textit{TESS} is less sensitive to small planets than \textit{Kepler} was, with \citet{Christ2018ObservationsFeatures} predicting that only 277 \textit{Kepler} planets have a >50\% likelihood of being recovered by \textit{TESS} to a 3$\sigma$ level in the wider-extent 30-min cadence data.

This issue is well illustrated in Figure \ref{fig:phot_comparison}, showing the comparison between \textit{Kepler} and \textit{TESS} data for a large Jupiter-sized planet (HAT-P-7 b; \citet{Pal2008HAT-P-7b:Field}) and a small Super-Earth sized planet (Kepler-21 b; \citet{Howell2012Kepler-21b:179070}). While the planetary signal for both systems is clear in the \textit{Kepler} data, the smaller planet is entirely indiscernible in the 2min \textit{TESS} light-curve. This perhaps helps to explain why only seven known \textit{Kepler} planets reobserved with \textit{TESS} 2min cadence were identified as \textit{TESS} Objects of Interest (TOIs).\footnote{https://exofop.ipac.caltech.edu/tess/view\_toi.php} 

For the purpose of this study we focus on all \textit{Kepler} confirmed planets and candidates with transits visible in the \textit{TESS} 2min data, including many that were missed as \textit{TESS} TOIs.

\subsection{Overall Target Selection} \label{target_selection}

In order to find all \textit{Kepler} planets and candidates which received 2min \textit{TESS} data, the 962 \textit{Kepler} stars with short-cadence data were cross-matched with all transiting planets and \textit{Kepler} planet candidates from the NASA Exoplanet Archive.\footnote{https:\/\/exoplanetarchive.ipac.caltech.edu [Accessed 12 August 2020]} This revealed a population of 49 \textit{Kepler} planetary host stars (harbouring 93 transiting planets) and 28 planet candidate systems which were re-observed by \textit{TESS} in its short-candence mode. However, due to the comparatively short (28-day) \textit{TESS} observation window compared to some long-period \textit{Kepler} planets, the expected transit times were missed for 19 planets and 8 planet candidates, making them unusable for ephemeris updates. Note that the planetary system of Kepler-34 (TIC 164457525) was also removed, as its planet is not expected to transit again until Nov 18 2066 due to its complex circumbinary orbit \citep{Welsh2015KEPLERPLANET,Martin2017CircumbinaryGo}. 

The \textit{TESS} 2min light-curves for the remaining 73 transiting planets and 20 planet candidates were searched for new transits both by eye and systematically with a box-least-squares search \citep{Kovacs2002ATransits,Hartman2016VARTOOLS:Data}, as implemented as the \texttt{BoxLeastSquares} function in the \texttt{astropy} python package \citep{exoplanet:astropy13,AstropyCollaboration2018ThePackage}. Because of the reduced sensitivity of the \textit{TESS} instrument, many of the smallest \textit{Kepler} planets were indiscernible from noise in the \textit{TESS} data alone, so these systems were also cut from the analysis. Overall there remained 22 planets (in 18 planetary systems) and 4 planet candidates where the ephemerides could be reasonably improved using the \textit{TESS} 2min data. It is worth noting however that for many of these only a single transit was observed. Table \ref{tab:target_selection} gives an overview of the final systems which have been updated in this study, while a system by system summary of this target selection is included for each planet host in Table \ref{tab:reobserved_table}.

\begin{table}
	\centering
	\caption{Overview of final Kepler planetary and candidate systems for which ephemerides were updated in this study. A TTV flag of '1' denotes any systems where unusual/non-linear behaviour was observed in the O-C diagrams constructed in this work.}
	\label{tab:target_selection}
	Planets\\
	\resizebox{\columnwidth}{!}{%
	\begin{tabular}{cccc} 
		\hline
		System Name & Planets updated & TTV flag & Discovery Paper\\
		\hline
		Kepler-2/HAT-P-7 & b & 1 & \citet{Pal2008HAT-P-7b:Field}\\
		Kepler-10 & c & 1 & \citet{Fressin2011Kepler-10System}\\
		Kepler-13/KOI-13 & b & 0 & \citet{Borucki2011CharacteristicsData}\\
		Kepler-14 & b & 0 & \citet{Buchhave2011Kepler-14b:Binary}\\
		Kepler-18 & d & 1 & \citet{Cochran2011Kepler-18bMeasurements}\\
		Kepler-25 & b, c & 1 & \citet{Steffen2012TransitVariations}\\
		Kepler-51 & b & 1 & \cite{Steffen2013TransitStability}\\
		Kepler-51 & d & 1 & \cite{Masuda2014VeryEvent}\\
		Kepler-63 & b & 0 & \cite{Sanchis-Ojeda2013Kepler-63b:Star}\\
		Kepler-68 & b & 0 & \cite{Gilliland2013Kepler-68:Giants}\\
		Kepler-89/KOI-94 & d & 1 & \citet{Weiss2013TheFlux}\\
		Kepler-96 & b & 0 & \citet{Marcy2014MassesPlanets}\\
		Kepler-289 & c & 1 & \citet{Rowe2014ValidationSystems}\\
		Kepler-396 & b, c & 1 & \citet{Xie2014TransitSystems}\\
		Kepler-411 & c & 1 & \citet{Morton2016FalsePositives}\\
		Kepler-411 & d & 1 & \citet{Sun2019Kepler-411:Star}\\
		Kepler-412 & b & 1 & \citet{Deleuil2014SOPHIEJupiter}\\
		Kepler-448/KOI-12 & b & 1 & \citet{Bourrier2015SOPHIERotator}\\
		Kepler-538 & b & 0 & \citet{Morton2016FalsePositives}\\
		Kepler-1517 & b & 0 & \citet{Morton2016FalsePositives}\\
		\hline
	\end{tabular}%
	}\\
	\vspace{5mm}
	Candidates\\
		\begin{tabular}{ccc} 
		\hline
		System Name & Candidates updated & TTV flag \\
		\hline
		KIC 7199397 & K00075.01 & 1 \\
		KIC 8554498 & K00005.01 & 0 \\
		KIC 9955262 & K00076.01 & 1 \\
		KIC 9418619 & K06068.01 & 0 \\
		\hline
	\end{tabular}
\end{table}

\section{Methods} \label{methods}


\subsection{Assembling Archival System Parameters} \label{assembling_priors}

As a starting point for the models, planetary and stellar parameters were retrieved from the NASA Exoplanet Archive for all objects. In cases where the data highlighted on the exoplanet archive was out of date, these paramaters were updated to reflect the most recent literature. In order to collect the most recent radii information, the entire sample was cross-matched with  \citet{Berger2018Revised2}'s revised radii of \textit{Kepler} stars and planets based on \textit{Gaia} DR2 data \citep{Lindegren2018GaiaSolution,GaiaCollaboration2018GaiaProperties}. This list was then cross-matched with the updated linear ephemerides found by \citet{Gajdos2019TransitExoplanets}, who used the entire length of \textit{Kepler} data (Q1-17) to update ephemerides for 1977 exoplanets. As a first check for significant transit timing variations, the TTV flag was checked on the Exoplanet Archive, and all systems cross-matched with TTV data from surveys completed by \citet{Holczer2016SET} and \citet{Gajdos2019TransitExoplanets}.

Similar to the \textit{Kepler} planets, original data for the \textit{Kepler} planet candidates was collected from the \textit{Kepler} Objects of Interest (KOI) list on the Exoplanet Archive, taking care to disregard any KOIs previously downgraded to false positives. These data were then cross-matched with data from the most recent TESS Input Catalog \citep[TICv8,][]{Stassun2019TheList} and \citet{Berger2018Revised2}'s revised radii of \textit{Kepler} stars and planets to obtain up to date information for the stellar hosts.

\subsection{Ephemeris updates and construction of O-C diagrams} \label{ephem_update}

The methods used to determine individual transit times and update linear ephemerides are based on those used by \citet{Gajdos2019TransitExoplanets} and \citet{Holczer2016SET} to analyse the full Kepler Q1-17 data. As a first step, transit epochs for planets/candidates other than the one being currently analysed were masked from the light-curve. In some cases with large TTVs (such as Kepler-396 b - \citet{Xie2014TransitSystems}), the masking window was widened slightly to catch all of the transits. Because of the differing dilution characteristics and data cadence of the \textit{Kepler} and \textit{TESS} instruments, the \textit{Kepler} transits were found first and then the results from this analysis were used to inform the initial \textit{TESS} model. 

To achieve consistent times for the two satellites, the time data for both datasets was converted to Barycentric Julian Date using the following conversions:

\begin{itemize}
    \item $Kepler_{BJD} = KBJD + 2454833$ days
    \item $TESS_{BJD} = BTJD + 2457000$ days
\end{itemize}

\begin{figure}    \includegraphics[width=\columnwidth]{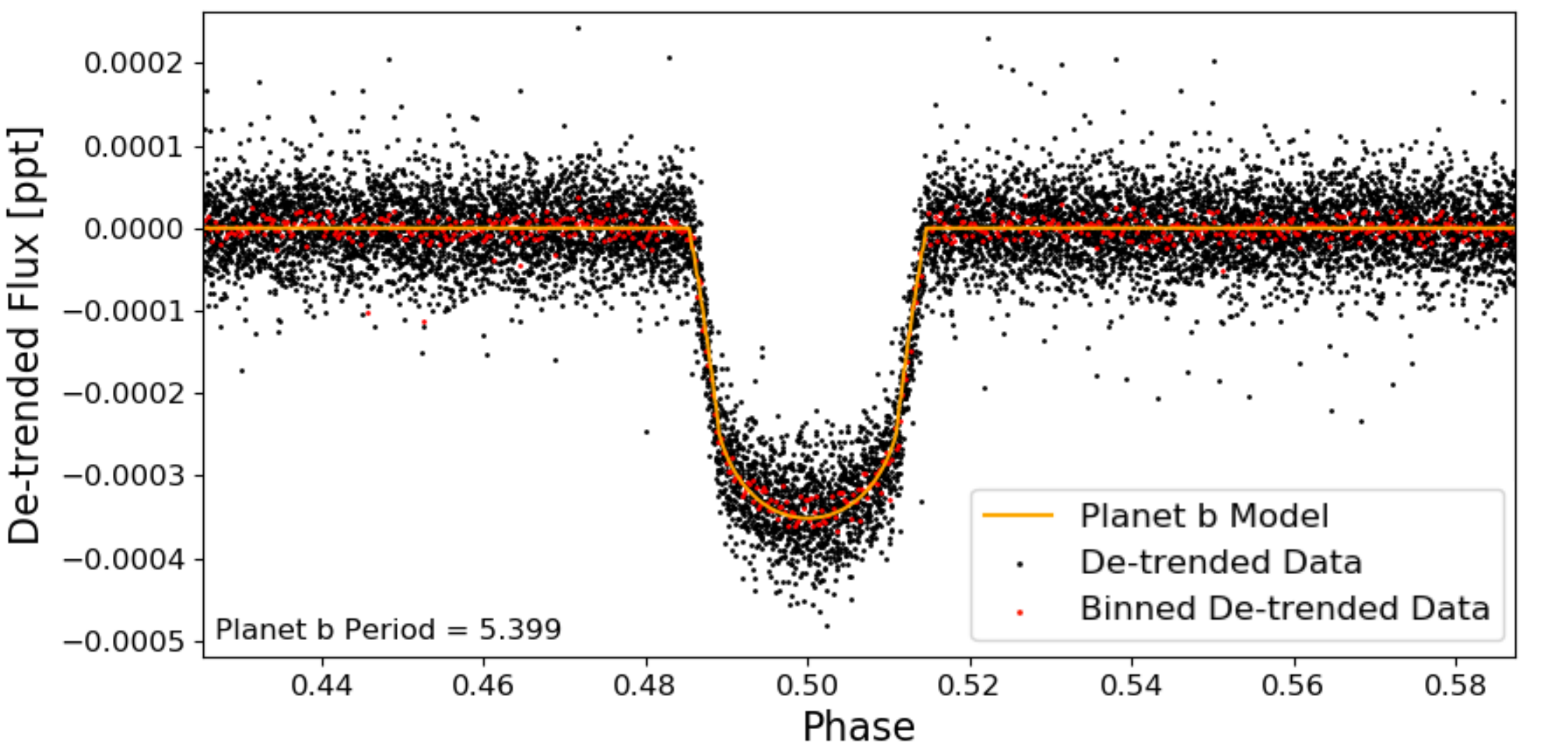}
    \caption{Stacked transit curve for Kepler-68 b overplotted with final \texttt{batman} model.
    }
     \label{fig:stacked_fig}
 \end{figure}

For the first step in the \textit{Kepler} analysis, a \citet{Mandel2002AnalyticSearches} transit model was constructed for the object of interest using \citet{Kreidberg2015Batman:Python}'s \texttt{batman} software, based off the archival parameters assembled in Section \ref{assembling_priors}. These models were used to search for the observed transit time by minimising the chi-squared statistic in a grid of 1min resolution around the expected transit. This method, similar to that used by \citet{Holczer2016SET}, ensured that the approximate initial transit times could be obtained automatically even when there were significant TTVs present. These 1min resolution transit times were used to construct a 'stacked' light-curve for each planet/candidate by aligning the obtained transit centres, similar to the method used by \citet{Gajdos2017Transit-timingKepler-410Ab,Gajdos2019TransitExoplanets} (see Figure \ref{fig:stacked_fig}). This stacked transit was compared to the transit curve obtained simply by folding the \textit{Kepler} data by the known planetary period, with the cleanest transit curve taken forward for further analysis. The dual approach ensured that accurate transit curves could be obtained for targets both with and without transit timing variations. The chosen transit curve was fitted with a new \texttt{batman} model using three iterations of the in-built \texttt{optimize} routine within \texttt{exoplanet} \citep{exoplanet:exoplanet}. Because the individual transit times here are considered more important than the overall system parameters, the planetary radius and inclination were allowed to vary slightly in order to fit the phase-curve most effectively. Note that both parameters were modelled as normal distributions with means equal to value reported in literature (or $90^o$ for systems without published inclinations). The mean value outside each transit was also allowed to vary in case it was skewed by outliers or detrending artifacts. An example of the final transit model can be seen for Kepler-68 b \citep{Gilliland2013Kepler-68:Giants} in Figure \ref{fig:stacked_fig}. 

The generated transit model was used to find precise transit times for individual transits in the \textit{Kepler} data. This was achieved by cutting out short-duration windows around each expected transit (2 days either side of the transit, unless the planet period was $\leq$2 days) and finding the best-fit of the transit model within that interval. For consistency only the transit time and mean flux were allowed to vary in this step, both of which were set as wide Normal distributions centred on the values found from the chi-squared fit. To begin with, three iterations of \texttt{exoplanet}'s inbuilt \texttt{optimize} routine were used to hone in on the true transit times, before PyMC3 \citep{exoplanet:pymc3} was used to fit the final planet model to each individual \textit{Kepler} transit. This was achieved using a two-chain Markov-Chain Monte Carlo (MCMC) analysis with 1000 tuning steps and 5000 draws for each object. Longer chains were tested briefly, but increasing the length further for these relatively simple fits was not found to change the results significantly. Convergence was assessed using the Gelman-Rubin statistic and a visual examination of the trace. Statistically significant values for the final time and error for each individual transit were then found from the mean and standard deviation of the MCMC trace.

Because of the dearth of \textit{TESS} transits, broadly the same model built from the \textit{Kepler} data was used to find the times of the individual \textit{TESS} transits, however the depth of the transit was adjusted according to the local dilution characteristics of the \textit{TESS} environments. Once again the model was fit to each transit in turn (in a 4-day data interval centred on the expected transit time) using three iterations of \texttt{exoplanet}'s \texttt{optimize} function and a \texttt{pyMC3} MCMC analysis with 10000 tuning steps and 10000 draws in order to gain statistically significant values and errors for each mid-transit time.

\begin{figure*}
    \includegraphics[width=\textwidth]{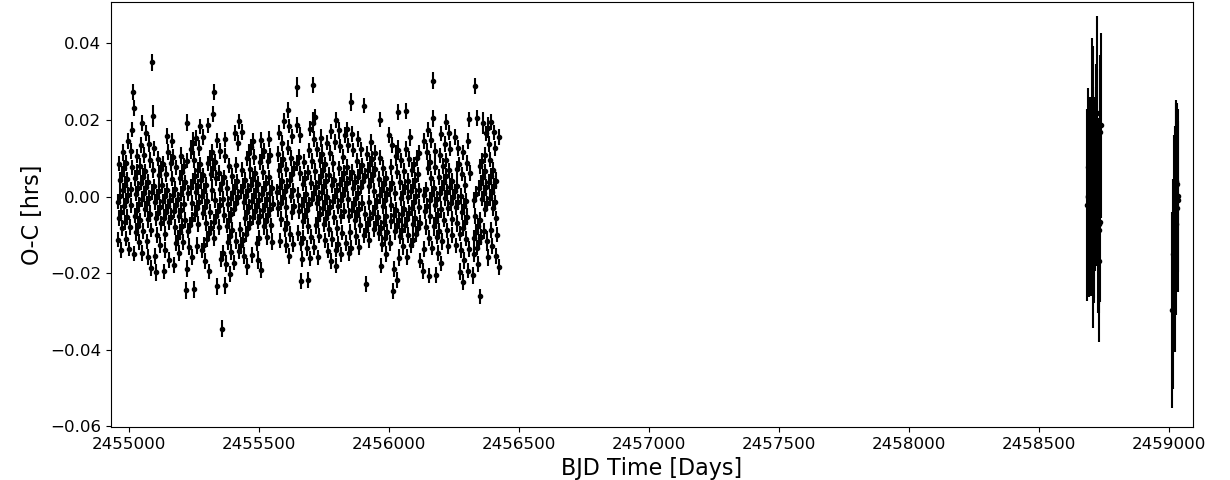}
    \caption{Example O-C plot for Kepler-13/KOI-13 b. K0I-13 b received \textit{TESS} 2min data in Sectors 14, 15 and 26, so represents one of the targets which received the most new data from \textit{TESS}'s primary mission.}
     \label{fig:o-c_example}
 \end{figure*}

After the time of each individual \textit{TESS} and \textit{Kepler} transit was determined, the linear ephemeris based on the combined datasets was calculated. An uncertainty-based requirement of $\sigma_{transit} < 0.2$hrs was found to remove the majority of the questionable transits prior to further analysis (typically those which fell in data gaps or were incomplete), however the remainder were viewed by eye to catch any other transits which clearly had been fitted incorrectly. The remaining data points were fit using three iterations of a weighted linear least-squares fit while varying the initial transit time, $T_0$, and period $P$. The resulting ephemerides were used to generate an Observed-Calculated (O-C) plot for each object, providing a visual check for odd behaviour, significant TTV signals or significant outliers. An example O-C plot is shown in Figure \ref{fig:o-c_example} for KOI-13 b. Some of the most interesting O-C diagrams are discussed in the results below (§\ref{results}).

Because of the long data gap between the \textit{Kepler} and \textit{TESS} observation windows, one slight concern was the introduction of cycle count errors, i.e. miscalculating the number of transits between the \textit{Kepler} and \textit{TESS} transits. To check whether this was significant, the archival error in period derived from the \textit{Kepler} data alone was multiplied by the number of missing transit cycles between the final \textit{Kepler} transit and first \textit{TESS} transit. In all cases this was found to result in a value at least two orders of magnitude smaller than a whole transit cycle, hence the ephemerides derived in this work are considered free of cycle count errors.

For any objects which did not display significant TTVs in their O-C diagrams (labelled with '0' in the TTV flag column of Table \ref{tab:target_selection}), an additional simultaneous \textit{Kepler/TESS} MCMC fit was carried out on the entire light-curve in order to narrow down the precision for these planets/candidates further. This was achieved by fitting the model based on the stacked transit parameters to the entire dataset and allowing t0, the mean out-of-transit value and the planetary period to vary. Once again, 1000 tuning steps and 5000 draws were used in this MCMC analysis, using pyMC3 \citep{exoplanet:pymc3} within \texttt{exoplanet} \citep{exoplanet:exoplanet}, following three \texttt{optimize} iterations. 


\subsection{TTV analysis} \label{TTV_analysis}

The updated systems in this work included a number of \textit{Kepler} multi-planet systems known to exhibit significant TTVs, namely Kepler-18, Kepler-25, Kepler-51, Kepler-89, and Kepler-396 \citep{Cochran2011Kepler-18bMeasurements, Steffen2012TransitVariations,Steffen2013TransitStability,Masuda2014VeryEvent,Weiss2013TheFlux,Xie2014TransitSystems}. Most of these systems were independently highlighted as worthwhile systems for further TTV analysis by \citet{Goldberg2018ProspectsObservations} and/or \citet{Jontof-Hutter2021FollowingCharacterization}. In an attempt to update the TTV masses for planets in these systems, additional TTV fitting was carried out. Midtransit times for each planet in a system were simulated using \texttt{TTVFast}, a symplectic integrator for computing transit times given a set of planet masses and orbital parameters \citep{Deck2014TTVFast:Inversion}. With these simulations, an MCMC analysis explored the parameter space to find the best-fit masses and orbits describing the observed transit times, both before and after the addition of \textit{TESS} datapoints.

Each transiting planet in a system was fit for mass, $\mu_{i}$ (in units of solar masses), orbital period, $P_{i}$, orbital eccentricity and argument of pericenter (via eccentricity vectors $h_{i} = \sqrt{e_{i}}\cos{\omega_{i}}$ and $k_{i} = \sqrt{e_{i}}\sin{\omega_{i}}$), and time of first transit, $T_{i}$, where $i = 1, 2, ..., N$ and $N$ is the number of planets. Coplanar orbits were assumed. Guess parameters were estimated using a Levenberg-Marquardt least-squares algorithm, and datapoints more than 4$\sigma$ from this initial best-fit solution were marked as outliers and removed from the data. MCMC analysis was then performed using the affine invariant ensemble sampler \texttt{emcee} \citep{Foreman-Mackey2012Emcee:Hammer}, with 100 walkers initialised in a tight ball around the guess parameters. Each walker was run for 200,000 steps, and the first 50,000 steps were discarded as burn-in.

Following the procedure of \citet{Hadden2016NumericalVariations} and \citet{Hadden2017KeplerAnalysis}, both ``default'' and ``highmass'' priors were considered, recognising that TTV fits are often challenged by mass-eccentricity degeneracies. Under the default run, mass was assigned a logarithmic prior

\begin{equation}
\begin{split}
    p(\mu) & \propto
    \begin{cases}
    (\mu + \mu_{0}) & \mu \geq 0, \\
    0 & \text{otherwise} \\
    \end{cases}
\end{split}
\end{equation}

\noindent with $\mu_{0} = 3\times10^{-7}$ to prevent divergence at $\mu \rightarrow 0$, while eccentricity was assigned a uniform prior

\begin{equation}
\begin{split}
    p(h, k) & \propto
    \begin{cases}
    (h^{2} + k^{2})^{-1/2} & (h^{2} + k^{2})^{1/2} < 0.9, \\
    0 & \text{otherwise} \\
    \end{cases}
\end{split}
\end{equation}

\noindent with an upper cutoff at 0.9 to avoid the need for extremely small time steps in the \texttt{TTVFast} integrations. For the high-mass run, mass was assigned a uniform prior

\begin{equation}
\begin{split}
    p(\mu) & \propto
    \begin{cases}
    \text{constant}, \mu \geq 0, \\
    0 & \text{otherwise} \\
    \end{cases}
\end{split}
\end{equation}

\noindent while eccentricity was assigned a logarithmic prior

\begin{equation}
\begin{split}
    p(h, k) & \propto
    \begin{cases}
    (h^{2} + k^{2})^{-1/2}(\sqrt{h^{2} + k^{2}} + e_{0})^{-1} & (h^{2} + k^{2})^{1/2} < 0.9, \\
    0 & \text{otherwise} \\
    \end{cases}
\end{split}
\end{equation}

\noindent with $e_{0} = 10^{-3}$ to prevent divergence at $e \rightarrow 0$. In both runs, period and initial transit time were assigned uniform priors.

\section{Results} \label{results}

\subsection{Known Planets}

All ephemerides updated in this study are presented in Table \ref{tab:ephem_table}, with known planets in the top section of the table and the four planet candidates presented underneath. Here the overall results for the known planets are presented, with a few of the most interesting systems investigated in more depth.

Overall good agreement was found for all updated planet ephemerides when compared to those from the \textit{Kepler} alone. With one exception (Kepler-51 d - \citet{Masuda2014VeryEvent}), all new periods were within 3$\sigma$ of their archival values, despite the very precise measurement errors. Any differences between new and archival $T_0$s can easily be explained by different individual transits being chosen as the zero point. The difference in period for Kepler-51 d is perhaps not surprising given the known transit timing variations of this planet coupled with its long period (130.2 days). Indeed, only nine reasonable transits were captured in the \textit{Kepler} data, and only a single transit in the data from  \textit{TESS}'s primary mission. Because of the >8 year gap between \textit{Kepler} and \textit{TESS} observations and the inclusion of the new \textit{TESS} data increasing the available number of data-points by 11\%, the new ephemeris found in this study is favoured. The transit timing variations for this planet are plotted in Figure \ref{fig:Kepler-51d} and explored further in Section \ref{ttv_results} below.

\begin{figure}
    \includegraphics[width=\columnwidth]{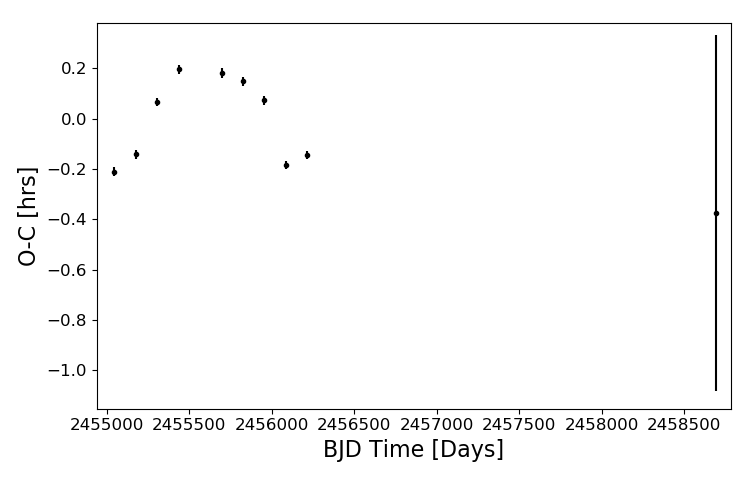}
    \caption{O-C plot for Kepler-51 d, showing evidence of transit timing variations after a linear ephemeris fit.}
     \label{fig:Kepler-51d}
\end{figure}

\begin{figure*}
    \includegraphics[width=\textwidth]{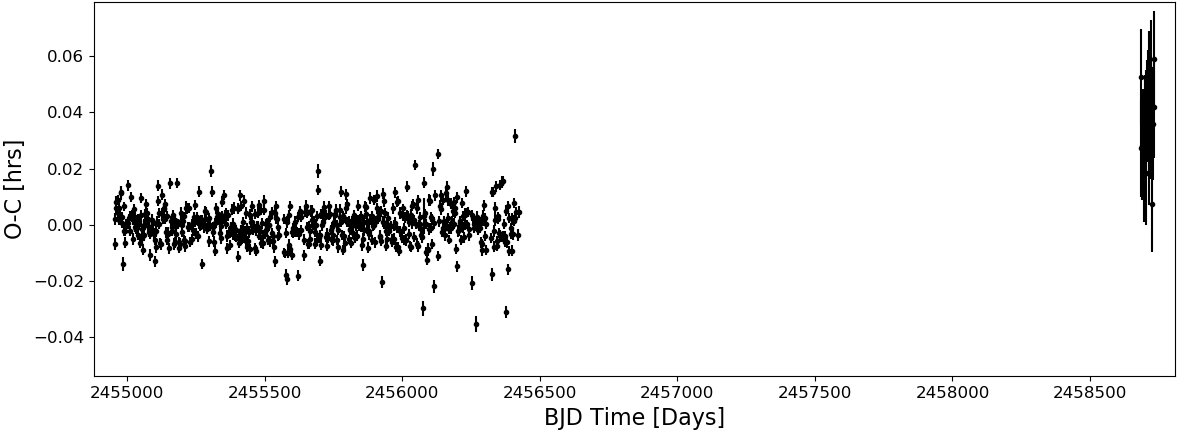}
    \caption{O-C plot for HAT-P-7b, showing notable offset between \textit{Kepler} (BJD < 24565000) and \textit{TESS} (BJD > 2458500) datasets.}
     \label{fig:HAT_P_7b}
\end{figure*} 

The overall precision for each ephemeris was found similar to those derived by \citet{Gajdos2019TransitExoplanets}, likely because both their study and this one used the entirety of the \textit{Kepler} data to define the planet ephemerides. The new \textit{TESS} data was thus most useful for improving the precision for planets whose most recent ephemerides came from different sources to \citet{Gajdos2019TransitExoplanets}, namely Kepler-2/HAT-P-7 b, Kepler-13 b, Kepler-18 d, Kepler-289 c, Kepler-396 b \& c, Kepler-411 c \& d and Kepler-538 b \citep[][see Table \ref{tab:ephem_table} for most recent ephemeris references]{Pal2008HAT-P-7b:Field,Borucki2011CharacteristicsData,Cochran2011Kepler-18bMeasurements,Rowe2014ValidationSystems,Xie2014TransitSystems,Morton2016FalsePositives,Sun2019Kepler-411:Star}. It should be noted that due to the typically shallower transits recovered in the \textit{TESS} data, the uncertainties in each \textit{TESS} transit are considerably higher than those of each \textit{Kepler} transit, which helps to explain why the new \textit{TESS} data is less constraining than initially expected.

Of the Kepler planets studied in this work, only KOI-13/Kepler-13 b (\citet{Borucki2010KeplerResults}, Figure \ref{fig:o-c_example}) has previously been updated with the newly available \textit{TESS} data. This is presented by \citet{Szabo2020TheKepler-13Ab}, who find a period of 1.76358760 $\pm$ 0.00000003 days, and a $T_0$ of 2455101.707254 $\pm$ 0.000012 BJD, in clear agreement with the values obtained in this work (see Table \ref{tab:ephem_table}). Similar to \cite{Szabo2020TheKepler-13Ab}, no evidence of TTVs or tidal decay of this Hot Jupiter was seen in this study. It should be noted that though the eventual ephemeris reported for Kepler-13b in this work was drawn from a simultaneous \textit{TESS/Kepler} fit of the entire light-curve, the ephemeris obtained using the initial TTV-focussed (transit-by-transit) fit was also in clear agreement, with a value of $T_0$ = 2454955.32946735 $\pm$ 0.000006153 and Period = 1.76358760059 $\pm$ 0.00000001281. This system thus provides a useful test-case for the overall methods used in the wider analysis carried out in this work.

One planetary O-C diagram of particular note in this work is that for Kepler-2b/HAT-P-7b, a bright (Gmag = 10.365), well-studied Hot Jupiter (2.0 $R_J$) originally discovered by \citet{Pal2008HAT-P-7b:Field}. This was previously identified as being a prime target for testing orbital decay by \citet{Christ2018ObservationsFeatures}. As can be clearly seen in Figure \ref{fig:HAT_P_7b}, a slight (2.5min) offset was found to exist between the original \textit{Kepler} data zero-point and new \textit{TESS} data. While extending the errors of the \textit{TESS} data points to two or three standard deviations would put them in agreement with the general \textit{Kepler} trend, the observed shift is still considered odd given that the excellent timing agreement for similar Hot Jupiter systems such as KOI-13/Kepler-13b and Kepler-412b. In an attempt to solve this discrepancy, additional transits for HAT-P-7 b were sought from the Exoplanet Transit Database\footnote{http://var2.astro.cz/ETD/etd.php?STARNAME=HAT-P-7\&PLANET=b} \citep{Poddany2010ExoplanetTransits}, however given the even larger timing uncertainties from the ground based data, no trend was evident. It should be noted though that the original discovery t0 \citep{Pal2008HAT-P-7b:Field} also aligns with a slight timing offset of 2.5min above this \textit{Kepler} zero-point. It is hoped that further data from space telescopes (such as when \textit{TESS} returns to the \textit{Kepler} field in year four of its mission) will help to test whether this shift is real.
 
\begin{figure}
    \includegraphics[width=\columnwidth]{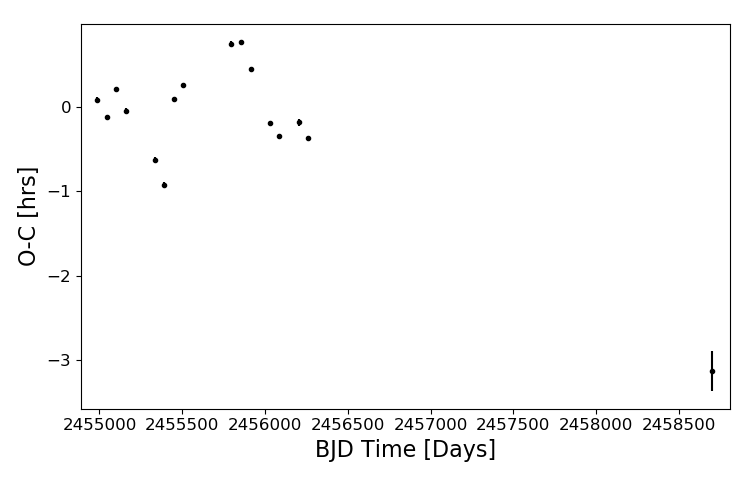}
    \caption{O-C plot for Kepler-411d, illustrating the ~3day shift between the Kepler data and single \textit{TESS} datapoint.}
     \label{fig:Kepler_411d}
\end{figure}

Another particularly interesting system updated here is Kepler-411 d (Figure \ref{fig:Kepler_411d}). While only one transit was caught in the \textit{TESS} data (due to the relatively long 58d period), this occurred over 3 hours before the expected epoch based on ephemerides in literature. Previous to this new \textit{TESS} data, Kepler-411 d had transit timing variations not explained by the inner planets in the system (Kepler-411 b \& c), with amplitudes of approximately one hour. \citet{Sun2019Kepler-411:Star} analysed these same TTVs (see Figure 8 from \citet{Sun2019Kepler-411:Star}) and found that they were best modelled by considering an extra non-transiting planet (31.5 day period) in the Kepler-411 system, Kepler-411 e. However, the new \textit{TESS} data suggests that the TTV amplitude may be much larger than previously thought, so more data (both photometric and spectroscopic) is recommended in order to constrain the parameters of this planetary system. Similarly large amplitude O-C variations were seen for Kepler-396c, so it is imperative that transit timing variations are taken into account when forecasting future observations in the Kepler-396 system (Figure \ref{fig:Kepler-396_system}).

Other than those systems discussed above, the confirmed planetary systems which received updated linear ephemerides in this work had transits that aligned well with the previous \textit{Kepler} observations and thus should be trustworthy for future observations. However, some care should be taken when forecasting individual transits within these systems, as approximately half of the updated systems exhibited non-linear behaviour in their O-C diagrams.

\subsection{Planet Candidates}

\begin{figure}
    \includegraphics[width=\columnwidth]{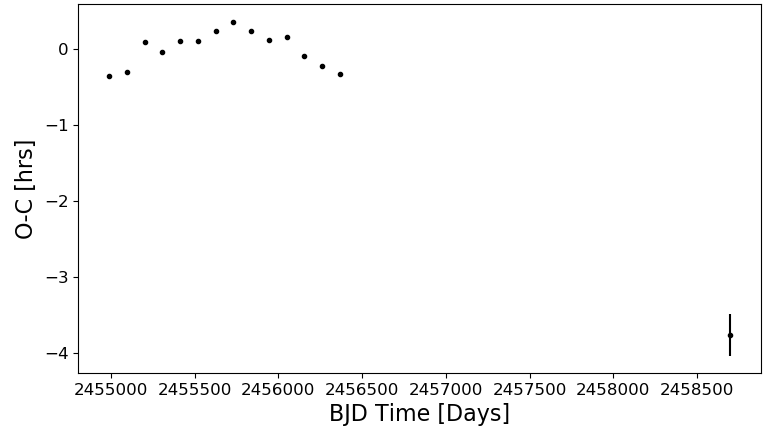}
    \caption{O-C plot for \textit{Kepler} candidate K00075.01 after fitting the linear ephemeris derived in this work. A non-linear ephemeris is considered more appropriate for this object, however is left for the future when more \textit{TESS} datapoints are available.}
     \label{fig:K00075_O-C}
\end{figure}

\begin{figure}
    \includegraphics[width=\columnwidth]{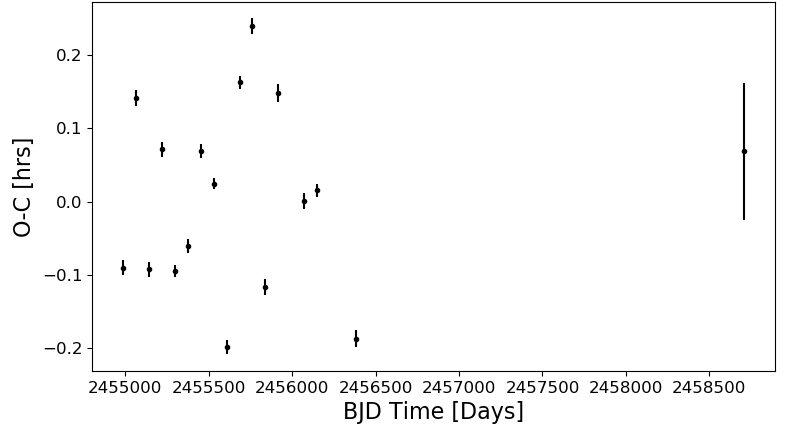}
    \caption{O-C plot for \textit{Kepler} candidate K00076.01. Some evidence of short-term transit-timing variations are apparent.}
     \label{fig:K00076_O-C}
\end{figure}

While issues such as transit epochs falling outside the \textit{TESS} observation window and the increased scatter of \textit{TESS} data precluded updating the ephemerides for the majority of the 28 \textit{Kepler} candidates, the \textit{TESS} data provided sufficient sensitivity to update four systems: KOI-5, KOI-75, KOI-76 and KOI-6068. The new ephemerides for the candidates in these systems can be found in the second part of Table \ref{tab:ephem_table}. The inclusion of the \textit{TESS} data significantly decreased the uncertainty in both the initial transit time ($T_0$) and the period  for these candidates. This new data was most helpful for the two longer-period candidates, K00075.01 (105.9d period) and K00076.01 (77.5d period), reducing their uncertainties by almost an order of magnitude despite only single transits being present in the comparatively short-duration \textit{TESS} observations. Care must be taken when using these new linear ephemerides however, as some evidence of transit timing variations can be seen in the O-C diagrams for both candidates (K00075.01 - Figure \ref{fig:K00075_O-C}; K00076.01 - Figure \ref{fig:K00076_O-C}). 

In the case of K00075.01, the O-C plot resulting from the weighted linear fit is clearly not linear, suggesting that it would be better fit with a non-linear ephemeris. \citet{VanEylen2019TheSystems} previously fit a sinusoidal model to the O-C plot for K00075.01, which suggested a period of 1892 days and a TTV amplitude of 22min. However, the new \textit{TESS} data point suggests a considerably larger amplitude is more appropriate. It is hoped that further transits of this candidate system from the extended \textit{TESS} mission will help to constrain the long-term TTV behaviour further.

Meanwhile, although the long-term trend of the K00076.01 system's ephemeris appears linear, there is evidence of shorter-term transit timing variations, with an amplitude of approximately 0.2hrs. This may be partially due to the stellar variability clear in the \textit{Kepler} light-curve, which has peak and trough-like features on a similar time-scale to the transit duration. Perhaps because of this variability, K00076.01 has been variously labelled both a candidate and false positive by different \textit{Kepler} KOI data releases, though most recently a candidate. More in-depth analysis of this individual system is recommended in future to test its planetary nature.

\begin{figure}
    \includegraphics[width=\columnwidth]{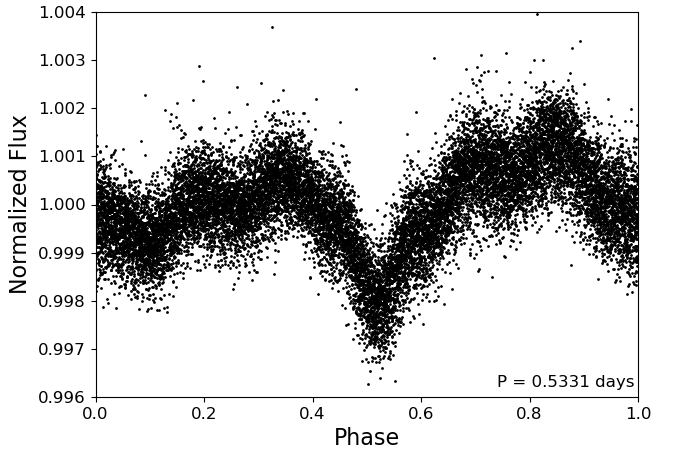}
    \caption{\textit{Kepler} candidate K00971.01 phase folded by KOI period. The faster cadence \textit{TESS} 2min data reveals significant evidence of stellar activity instead of a true planetary signal.}
     \label{fig:K00971_phase_fold}
\end{figure}

It should also be noted that while the O-C diagram for K06068.01 is flat and well-constrained, it has been labelled as a false positive in some past catalogs (e.g. q1\_q16\_koi \citep{Mullally2015PlanetaryMonths}, q1\_q17\_dr25\_koi \citep{Thompson2018Planetary25b}). Nonetheless, since the \textit{TESS} transit appears planet-like and the system was considered a candidate in the most recent cumulative and q1\_q17\_dr25\_sup\_koi KOI delivieries, it is still retained here. In contrast, candidate transits were also plausibly recovered for K00971.01, however eyeballing of this object in the higher-cadence \textit{TESS} data suggested that host star variability is a more likely hypothesis (see Figure \ref{fig:K00971_phase_fold}).

Thus while follow-up of \textit{Kepler} candidates with data from \textit{TESS} is somewhat hampered by the short observation windows and decreased sensitivity, this new source of data provides significant opportunities for improved characterisation of \textit{Kepler} candidates, both in transit timing and shape. This is particularly true for long-period \textit{Kepler} candidates, where relatively few data points exist in the \textit{Kepler} data alone. Furthermore, the shorter-cadence \textit{TESS} data is already revealing further details in the photometry to help to separate true planets/candidates from alternate false-positive scenarios.

\begin{table*}
	\centering
	\caption{Overview of all ephemerides updated in this work. Planet ephemerides are presented first, followed by the four planet candidates. Presented uncertainties are the 1$\sigma$ uncertainty for each value. n.b. 'Ref' refers to the reference each component of the archival ephemerides are drawn from, typically the most recent for each system: 1. \citet{Bonomo2017ThePlanets}; 2. \citet{Gajdos2019TransitExoplanets}; 3. \citet{Szabo2020TheKepler-13Ab} 4. \citet{Holczer2016SET}; 5. \citet{Su2020Mid-infraredSystems}; 6. \citet{Mayo2019AnStar}. All archival information for the four candidates was retrieved from the $q1\_q17\_dr25\_koi$ KOI data release \citep{Thompson2018Planetary25b}. A machine-readable copy of this table can be found in the online supplementary material.}
	\label{tab:ephem_table}
	Planets\\
	\resizebox{\textwidth}{!}{%
	\begin{tabular}{cccccc} 
		\hline
		Planet Name & Archival $T_0$ [BJD] & Archival Period [days] & Ref. &  Updated $T_0$ [BJD] & Updated Period [days]  \\ 
		\hline
		Kepler-2b & 2454954.357462 $\pm$ 0.000005 & 2.204737 $\pm$ 0.000017 & 1 & 2454954.3585572 $\pm$ 0.0000063 & 2.20473539167 $\pm$ 0.00000001654  \\
		Kepler-10c & 2455062.2665100 $\pm$ 0.0004297 & 45.29430079 $\pm$ 0.00003051 & 2 & 2454971.6772661 $\pm$ 0.0006847 & 45.29426146 $\pm$ 0.00003783\\
		Kepler-13b & 2455101.707254 $\pm$ 0.000012 & 1.76358760 $\pm$ 0.00000003 & 3 & 2454953.56595833 $\pm$ 0.00003827 & 1.76358750002 $\pm$ 0.00000002147\\
		Kepler-14b & 2454971.08821000$\pm$ 0.00006808 & 6.7901211029 $\pm$ 0.0000005613 & 2 & 2454957.50815829 $\pm$ 0.00008051 & 6.7901236131 $\pm$ 0.0000003985\\
		Kepler-18d & 2454961.155156 $\pm$ 0.000661 & 14.85891873 $\pm$ 0.00000074 & 4 & 2454961.1542075 $\pm$ 0.0002423 & 14.858908757 $\pm$ 0.0000004364\\
		Kepler-25b & 2455004.7108100 $\pm$ 0.0001007 & 6.2385326915 $\pm$ 0.0000007349 & 2 & 2454954.7979391 $\pm$ 0.0002168 & 6.2385347882 $\pm$ 0.0000001619\\
		Kepler-25c & 2455011.52792000 $\pm$ 0.00007383 & 12.720374906 $\pm$ 0.000001158 & 2 & 2454960.6467450 $\pm$ 0.0001144 & 12.720370495 $\pm$ 0.000001703\\
		Kepler-51b & 2455714.5917200 $\pm$ 0.0001741 & 45.15530956 $\pm$ 0.00001897 & 2 & 2454992.10682769 $\pm$ 0.0003851 & 45.15529233 $\pm$ 0.00002211\\
		Kepler-51d & 2455695.9210000 $\pm$ 0.0002442 & 130.17662541 $\pm$ 0.00007274 & 2 & 2455045.0339014 $\pm$ 0.0004519 & 130.17784455 $\pm$ 0.00008134\\
		Kepler-63b & 2455010.84340000 $\pm$ 0.00002768 & 9.4341503479 $\pm$ 0.0000003339 & 2 & 2454954.23899519 $\pm$ 0.00003794 & 9.4341522797 $\pm$ 0.0000004323\\
		Kepler-68b &  2455006.85878000 $\pm$ 0.00007639 & 5.3987525913 $\pm$ 0.0000005231 & 2 & 2454958.2700925 $\pm$ 0.0009446 & 5.398752420 $\pm$ 0.000003113 \\
		Kepler-89d & 2454965.7417600 $\pm$ 0.0001014 & 22.342971172 $\pm$ 0.000002603 & 2 & 2454965.7413033 $\pm$ 0.0001169 & 22.342982351 $\pm$ 0.000002982\\
		Kepler-96b & 2455004.02020000 $\pm$ 0.00008997 & 16.238459306 $\pm$ 0.000001893 & 2 & 2454955.3013031 $\pm$ 0.0001767 & 16.238459945 $\pm$ 0.000005665\\
		Kepler-289c & 2455069.661672 $\pm$ 0.002848 & 125.86526539 $\pm$ 0.00000325 & 4 & 2455069.6605154 $\pm$ 0.0002487 & 125.86521071 $\pm$ 0.00004151\\
		Kepler-396b & 2454995.495267 $\pm$ 0.005940 & 42.99292187 $\pm$ 0.00000635 & 4 & 2454995.4942951 $\pm$ 0.0004076 & 42.99292140 $\pm$ 0.00002072\\
		Kepler-396c & 2455015.677038 $\pm$ 0.019460 & 88.51067812 $\pm$ 0.00002174 & 4 & 2455104.1858280 $\pm$ 0.0002890 & 88.51097554 $\pm$ 0.00003917\\
		Kepler-411c & 2454968.2224 $\pm$ 0.0002 & 7.834435 $\pm$ 0.000002 & 5 & 2454960.3876333 $\pm$ 0.0001131 & 7.834436247 $\pm$ 0.000001137\\
		Kepler-411d & 2454984.8484 $\pm$ 0.0061 & 58.02035 $\pm$ 0.00056 & 5 & 2454984.8478013 $\pm$ 0.0005473 & 58.02023116 $\pm$ 0.00004203\\
		Kepler-412b & 2454966.02102000 $\pm$ 0.00002357& 1.7208612825 $\pm$ 0.0000000491 & 2 & 2454966.02101665 $\pm$ 0.00002771 & 1.72086125797 $\pm$ 0.00000005710\\
		Kepler-448b & 2454979.59635000 $\pm$ 0.00002880 & 17.8552258437 $\pm$ 0.0000006438 & 2 & 2454961.74166158 $\pm$ 0.00002718 & 17.8552273080 $\pm$ 0.0000005863\\
		Kepler-538b & 2455044.6789 $\pm$ 0.0010 & 81.73778 $\pm$ 0.00013 & 6 & 2454962.9402449 $\pm$ 0.0007438 & 81.73797957 $\pm$ 0.00007330\\
		Kepler-1517b & 2454966.50342000 $\pm$ 0.00006786 & 5.5460843094 $\pm$ 0.0000004568 & 2 & 2454955.4112371 $\pm$ 0.0000758 & 5.5460845139 $\pm$ 0.0000005019\\
		\hline
	\end{tabular}%
	}\\
	\vspace{5mm}
	Planet Candidates\\
		\begin{tabular}{cccccc} 
		\hline
		Candidate Name & Archival $T_0$ [BJD] & Archival Period [days] & Updated $T_0$ [BJD] & Updated Period [days] \\
		\hline
		K00005.01 & 2454965.974086 $\pm$ 0.000148 & 4.780327581 $\pm$ 0.000000852 & 2454956.4131442 $\pm$ 0.0003056 & 4.7803297849 $\pm$ 0.0000006835\\
		K00075.01 & 2454989.97935 $\pm$ 0.00101 & 105.8817667 $\pm$ 0.0001312 & 2454989.9831552 $\pm$ 0.0003309 & 105.88145696 $\pm$ 0.00004402\\
		K00076.01 & 2454987.70734 $\pm$ 0.00230 & 77.4794704 $\pm$ 0.0002456 & 2454987.7051992 $\pm$ 0.0001855 & 77.47983018 $\pm$ 0.00001939\\
		K06068.01 & 2454967.4228253 $\pm$ 0.0000879 & 6.15025138 $\pm$ 0.00000070 & 2454955.12232977 $\pm$ 0.00003749 & 6.1502525063 $\pm$ 0.0000003761\\
		\hline
	\end{tabular}
\end{table*}

\subsection{TTV analysis} \label{ttv_results}

\begin{figure}
    \centering
    \includegraphics[width=\columnwidth]{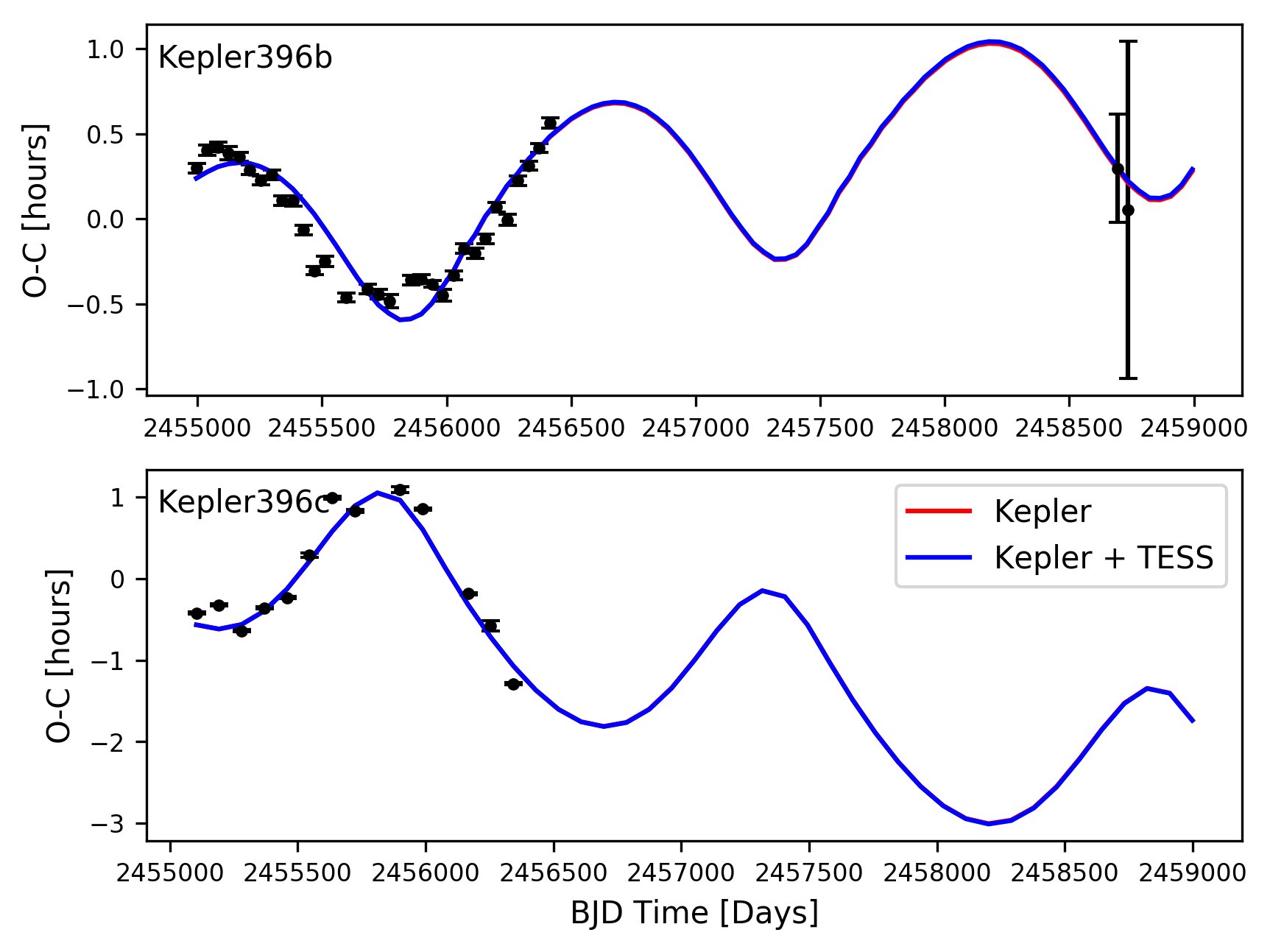}
    \caption{O-C diagrams for the Kepler-396 system (b: top; c: bottom) showing clear evidence of transit timing variations for each planet. The best-fit TTV results assuming a default prior, using only \textit{Kepler} data (red) and using the additional \textit{TESS} data for Kepler-396b (blue), are plotted. The solutions lie almost directly on top of one another.}
    \label{fig:Kepler-396_system}
\end{figure}

TTV fits were attempted for \textit{Kepler} multiplanet systems known to exhibit significant TTVs that were re-observed by \textit{TESS}. Five systems we identified had at least one planet with ephemerides recoverable from \textit{TESS} data: Kepler-18, Kepler-25, Kepler-51, Kepler-89, and Kepler-396. Inferred masses for the planets in these systems are shown in Table \ref{tab:ttv_table}, giving the median and $68.3\%$ credible region of the MCMC posteriors for both default and highmass runs. Figure \ref{fig:Kepler-396_system} demonstrates the TTV results for the two-planet Kepler-396 system using the default prior as an example.

While continued transit timing variations were clear in most systems, any differences between the mass results before and after the \textit{TESS} datapoints are well within error, and uncertainties on masses did not consistently decrease with the additional data. We conclude that the TTV fits did not improve with the \textit{TESS} datapoints, which we attribute to the large uncertainties associated with each new transit time, which were often larger than the TTV amplitudes themselves. Overall, the analysis is challenged by the faintness of the \textit{Kepler} targets, and the difficulty to see individual transits in the \textit{TESS} data.

\begin{table}
	\centering
	\caption{Overview of all TTV masses inferred in this work. Central values are the median of the MCMC posteriors, while lower and upper uncertainties are calculated from the 15.9th and 84.1th percentiles, representing the 68.3\% credible region.}
	\label{tab:ttv_table}
	Default prior\\
	\begin{tabular}{cccc}
		\hline
		Planet Name & Period [days] & Mass [$M_{\oplus}$] & Mass [$M_{\oplus}$] \\ 
		& & (\textit{Kepler}) & (\textit{Kepler} + \textit{TESS}) \\
		\hline
		Kepler-18c & 7.64 & $8.6_{-3.5}^{+5.4}$ & $8.4_{-3.3}^{+5.6}$\\
		Kepler-18d & 14.86 & $8.5_{-2.5}^{+2.5}$ & $8.4_{-2.3}^{+2.6}$ \\
		Kepler-25b & 6.24 & $1.0_{-0.5}^{+1.1}$ & $1.2_{-0.5}^{+1.5}$ \\
		Kepler-25c & 12.72 & $4.5_{-1.9}^{+3.2}$ & $5.0_{-1.9}^{+3.7}$\\
		Kepler-51b & 45.16 & $0.9_{-0.8}^{+1.3}$ & $1.1_{-0.9}^{+1.2}$ \\
		Kepler-51c & 85.32 & $2.8_{-0.4}^{+0.5}$ & $2.8_{-0.4}^{+0.5}$\\
		Kepler-51d & 130.18 & $4.4_{-0.9}^{+1.1}$ & $4.4_{-0.9}^{+1.0}$\\
		Kepler-89c & 10.42 & $5.3_{-1.5}^{+2.0}$ & $5.2_{-1.5}^{+2.1}$\\
		Kepler-89d & 22.34 & $39.4_{-8.1}^{+9.0}$& $39.3_{-9.0}^{+8.9}$\\
		Kepler-396b & 42.99 & $1.4_{-0.1}^{+0.1}$& $1.4_{-0.1}^{+0.1}$\\
		Kepler-396c & 88.51 & $1.1_{-0.1}^{+0.1}$ & $1.1_{-0.1}^{+0.1}$\\
	\end{tabular}\\
	\vspace{5mm}
	Highmass prior\\
	\begin{tabular}{cccc} 
		\hline
		Planet Name & Period [days] & Mass [$M_{\oplus}$] & Mass [$M_{\oplus}$] \\
		& & (\textit{Kepler}) & (\textit{Kepler} + \textit{TESS}) \\
		\hline
		Kepler-18c & 7.64 & $14.9_{-3.9}^{+3.8}$ & $15.1_{-3.7}^{+3.8}$\\
		Kepler-18d & 14.86 & $11.3_{-1.6}^{+1.3}$ & $11.4_{-1.5}^{+1.3}$\\
		Kepler-25b & 6.24 & $4.8_{-2.2}^{+3.8}$ & $5.0_{-2.2}^{+3.9}$\\
		Kepler-25c & 12.72 & $11.7_{-3.2}^{+2.8}$ & $11.9_{-3.0}^{+2.7}$\\
		Kepler-51b & 45.16 & $2.2_{-1.1}^{+1.3}$ & $2.3_{-1.1}^{+1.3}$\\
		Kepler-51c & 85.32 & $3.3_{-0.4}^{+0.4}$ & $3.4_{-0.4}^{+0.4}$\\
		Kepler-51d & 130.18 & $5.2_{-1.0}^{+1.0}$ & $5.2_{-1.0}^{+1.0}$\\
		Kepler-89c & 10.42 & $7.6_{-1.9}^{+2.5}$ & $7.5_{-1.9}^{+2.4}$ \\
		Kepler-89d & 22.34 & $42.6_{-7.3}^{+7.7}$& $42.1_{-7.4}^{+7.8}$\\
		Kepler-396b & 42.99 & $1.4_{-0.1}^{+0.1}$ & $1.4_{-0.1}^{+0.2}$ \\
		Kepler-396c & 88.51 & $1.1_{-0.1}^{+0.1}$ & $1.2_{-0.1}^{+0.2}$ \\
		\hline
	\end{tabular}\\
\end{table}

\section{Discussion} \label{discussion}

Precise and accurate ephemerides are crucial to the success and efficiency of future planet characterisation missions such as JWST. Without regular updates to their ephemerides, increasing timing uncertainties in known planets and planet candidates lead to increased observation costs and lost time, especially for longer-period systems.

In this work the ephemerides of 22 \textit{Kepler} planets and 4 planet candidates have been updated by analysing new transits in the 2min-cadence data from \textit{TESS}'s primary mission. However, the extent to which ephemeris updates were possible was less than was originally anticipated. The analysis carried out by \citet{Christ2018ObservationsFeatures} prior to the observations of the \textit{Kepler} field by \textit{TESS} was found to be excellent for determining recovery, however was perhaps a little optimistic for individual transit fitting. This was particularly the case for targets such as Kepler-10 b, Kepler-93 b, Kepler-138 c and Kepler-411 b \citep{Batalha2011KeplersKepler-10b,Marcy2014MassesPlanets,Kipping2014TheDwarfs,Wang2014InfluenceCandidates}, which were recovered when all transits were considered, but too shallow to allow individual transits to be fit reliably. For these objects (and other objects of similarly low signal to noise), an alternative method of ephemeris update is advised, perhaps by averaging all \textit{TESS} observations by sector/year.  

On a similar note, the null-result of the updated TTV analysis is an important, yet slightly disheartening one, as it suggests that updating inferred planet masses from TTVs with \textit{TESS} may be more challenging than the community hoped. Both the Kepler-396 and Kepler-51 systems were highlighted as high priority systems for improvement with \textit{TESS} by \citet{Goldberg2018ProspectsObservations}, yet neither was significantly improved by the data from the \textit{TESS} primary mission. The biggest challenge faced in updating these TTV masses is the faintness of most \textit{Kepler} systems with significant transit timing variations in the \textit{TESS} photometry, leading to very large uncertainties in individual transit times. When coupled with the fact that the systems analysed typically only had 1-3 transits in the \textit{TESS} data, the new data is not yet very constraining. It is hoped that this situation will improve as the \textit{TESS} extended mission continues. 

These results provide an interesting comparison of \textit{Kepler} and \textit{TESS} photometry in practice and illustrate the differences between optimisation of \textit{TESS} and \textit{Kepler} instruments. One thing that is immediately clear is how impressive the \textit{Kepler} satellite was for analysis of ephemerides and TTVs for fainter stars. These differences are particularly striking in Figure \ref{fig:phot_comparison}, clearly illustrating the difference in noise in the \textit{Kepler} and \textit{TESS} observations. The fact that the new \textit{TESS} data did not significantly change the uncertainties in the periods for those analysed by \citet{Gajdos2019TransitExoplanets} is testament both to their analysis and the very high quality of \textit{Kepler data}. For short-period \textit{Kepler} planets with many transits in the \textit{Kepler} data, it is perhaps unsurpsiring that only 1-3 sectors of \textit{TESS} data would change the already very precise ephemerides. This is especially the case for such dim stars as make up the bulk of the \textit{Kepler} field, since \textit{TESS} is deliberately designed to search for planets around brighter stars \citep{Rickeretal.2014TheSatellite}. Nonetheless, for those planets without \citet{Gajdos2019TransitExoplanets} ephemerides, \textit{TESS} has once again proven itself a very powerful tool for improving the precision of planet and candidate ephemerides, decreasing period uncertainty by orders of magnitude in cases such as HAT-P-7/Kepler-2b, Kepler-411d, Kepler-538b and the candidates K00075.01/K00076.01. 

One limitation of this work is that it focused only on \textit{Kepler} targets which received 2min data from the \textit{TESS} mission. While this data-set represents the shortest-cadence data so far released by the \textit{TESS} mission, it suffers from the choices made for which systems were put forward for 2min data collection. Given that the \textit{TESS} 2min catalog prioritises bright dwarf stars in its search for smaller planets \citep{Stassun2019TheList}, many of the dim stars in the \textit{Kepler} field did not receive 2min data. The analysis completed by \citet{Christ2018ObservationsFeatures} demonstrates the significant promise of extending this analysis to the \textit{TESS} Full Frame Images (FFIs), especially when the \textit{Kepler} field is revisited in extended mission. Such an analysis has the potential to increase the number of updated ephemerides from the approximately 25 systems updated here to hundreds of systems. This analysis is made significantly easier with the recent release of the \textit{TESS} SPOC \citep{Jenkins2016TheCenter,Caldwell2020TESSProducts} and Quick Look Pipeline \citep{Huang2020PhotometryImages} FFI light-curves on MAST. The planned 10min cadence data will also help considerably with this effort. 

However, given the challenges faced in this study due to the dimness of many \textit{Kepler} planets and candidates, it is imperative that we also find alternative methods/data-sources to maintain the ephemerides of dim or small \textit{Kepler} planets/candidates. Failing to do so risks letting many interesting systems fade into timing uncertainty oblivion.

\section{Summary}

In this work ephemerides have been updated for 22 \textit{Kepler} planets and 4 planet candidates using short-cadence data from the \textit{Transiting Exoplanet Survey Satellite}. 
This represents all \textit{Kepler} planets and candidates which so far have received \textit{TESS} 2min data with sufficient signal-to-noise to allow ephemeris updates to be carried out. The primary challenges to updating more systems were long period planets/candidates falling outside the \textit{TESS} observation window, and systems being too dim in the \textit{TESS} data. Because of the dimness of these objects, a purely photometric method was used to updated these ephemerides, using transits from \textit{Kepler} and \textit{TESS} data only. Transit times for individual transits were recovered by using an MCMC fit of a transit model based on stacked \textit{Kepler} transits. These transit times were then fit a weighted-linear fit in order to obtain the final ephemerides for each object. Any systems which did not appear to have significant transit timing variations were additionally fit with a simultaneous \textit{TESS/Kepler} fit to improve the epehmerides further. The resulting ephemerides were in good agreement with archival ephemerides and drastically reduced the uncertainty in period for Kepler-411d, Kepler-538b and the candidates K00075.01/K00076.01. Residuals to the linear ephemeris fits gave an important window into transit timing variations for these objects. TTV fits were attempted for five multi-planet systems known to exhibit significant TTVs (Kepler-18, Kepler-25, Kepler-51, Kepler-89, and Kepler-396), however these were challenged by the relative scarcity of reasonable \textit{TESS} transits and the dimness of these systems in \textit{TESS}. In the end there were no significant differences between the TTV masses before and after the inclusion of the \textit{TESS} data. More data is required (preferably of higher signal to noise) in order to constrain the TTV masses more effectively. Interesting TTV behaviour was observed also in the O-C diagrams for HAT-P-7b/Kepler-2b, Kepler-411d, K00075.01 and K00076.01, which warrants further analysis when new data becomes available. Thus while ephemeris updates and TTV analysis of \textit{Kepler} systems reobserved by \textit{TESS} proved less constraining than originally anticipated, \textit{TESS} has once again proved itself as an important follow-up instrument in addition to its primary planet-finding and asteroseismic aims. Overall, the ephemerides improved and updated in this study extend the life of these \textit{Kepler} systems, and improve prospects for future characterisation. Significant additional benefits are likely if the methods outlined in this paper are extended to the much larger sample of planets and candidates reobserved in the \textit{TESS} 30min cadence data, especially as \textit{TESS} returns to the \textit{Kepler} field in its extended mission. Care must be taken however to ensure that the community develops equipment and methods to maintain the ephemerides of dimmer \textit{Kepler} systems, as many of these are simply too dim to be followed up by \textit{TESS}.

\section*{Acknowledgements}

The authors would like to thank the anonymous referee for their comments which improved the quality and robustness of this paper. MPB wishes to acknowledge A Osborn for their helpful \texttt{exoplanet} tutorial.

This research made use of \texttt{exoplanet} \citep{exoplanet:exoplanet} and its
dependencies \citep{exoplanet:arviz, exoplanet:astropy13,
AstropyCollaboration2018ThePackage, exoplanet:exoplanet, exoplanet:kipping13,
exoplanet:luger18, exoplanet:pymc3, exoplanet:theano}. This paper includes data collected by the Kepler mission. Funding for the Kepler mission is provided
by the NASA Science Mission Directorate. This paper includes data collected by the \textit{TESS} mission. We acknowledge the use of public TOI Release data from pipelines at the \textit{TESS} Science Office and at the \textit{TESS} Science Processing Operations Center. Funding for the \textit{TESS} mission is provided by NASA’s Science Mission directorate. This research has made use of the Exoplanet Follow-up Observation Program website, which is operated by the California Institute of Technology, under contract with the National Aeronautics and Space Administration under the Exoplanet Exploration Program. \textit{TESS} and \textit{Kepler} data were obtained from the Mikulski Archive for Space Telescopes (MAST). STScI is operated by the Association of Universities for Research in Astronomy, Inc., under NASA contract NAS5-26555. Support for MAST is provided by the NASA Office of Space Science via grant NNX13AC07G and by other grants and contracts. 

This research has made use of the NASA Exoplanet Archive, which is operated by the California Institute of Technology, under contract with the National Aeronautics and Space Administration under the Exoplanet Exploration Program.\\
MPB acknowledges support from the University of Warwick via the Chancellor's International Scholarship. DJA acknowledges support from the STFC via an Ernest Rutherford Fellowship (ST/R00384X/1).

\section*{Data Availability}

The new ephemeris data generated in this article are available in the article and in its online supplementary material. The derived O-C transit timing data for individual planets analysed in this work will be shared on reasonable request to the corresponding author. All planet and candidate light-curves analysed in this work were drawn from public \textit{TESS} and \textit{Kepler} data hosted on the Mikulski Archive for Space Telescopes (MAST) Portal.\footnote{https://mast.stsci.edu/portal/Mashup/Clients/Mast/Portal.html} Additional parameters for the planet and candidate systems were drawn from the Exoplanet Archive,\footnote{https://exoplanetarchive.ipac.caltech.edu/} as discussed in the main text.


\bibliographystyle{mnras}
\bibliography{references} 



\appendix

\section{Overview of all \textit{Kepler} systems with \textit{TESS} 2min data. }

Table \ref{tab:reobserved_table} summarises all \textit{Kepler} systems with known transiting planets which received short-cadence (2min) data in \textit{TESS}'s Primary Mission. For each system all known planets are listed, along with whether their ephemerides were updated. For planets which were theoretically reobserved but not updated in this work, a brief explanation is supplied for their exclusion. Most commonly this is because their expected transits fell outside the eventual \textit{TESS} observation window, or they were shallow enough that any potential transits were lost in the increased noise of the \textit{TESS} data.
 
\begin{table*}
	\centering
	\caption{Overview of all \textit{Kepler} planetary systems which received \textit{TESS} short cadence data during \textit{TESS}'s Primary Mission. N.b. 'outside' = any planets whose expected transit epochs fell outside times of \textit{TESS} observations; 'shallow' = planets with transits which were sufficiently shallow for individual transits to be indiscernible from noise in the \textit{TESS} data.}
	\label{tab:reobserved_table}
	\begin{tabular}{cccc}
		\hline
		Kepler Name & TIC ID & Ephemeris Updated? & Reason for exclusion \\ 
		\hline
		Kepler-2 b & TIC 424865156 & \textbf{Yes} & -\\
		Kepler-9 b,c,d & TIC 120571842 & No & c outside; b,d shallow\\
		Kepler-10 b,c & TIC 377780790 & \textbf{Yes} - c, No - b & b shallow \\
		Kepler-11 b,c,d,e,f,g & TIC 169175503 & No & All shallow \\ 
		Kepler-13 b & TIC 158324245 & \textbf{Yes} & - \\
		Kepler-14 b & TIC 158561566 & \textbf{Yes} & - \\ 
		Kepler-16 b & TIC 299096355 & No & Outside\\
		Kepler-18 b,c,d & TIC 273690178 & \textbf{Yes} - d, No b,c & b,c shallow\\
		Kepler-21 b & TIC 121214185 & No & Shallow\\
		Kepler-25 b,c,d & TIC 120960812 & \textbf{Yes} - b,c; No - d & d outside\\
		Kepler-30 b,c,d & TIC 399794329 & No & b shallow; c,d outside\\
		Kepler-34 b & TIC 272369124 & No & Outside\\
		Kepler-35 b & TIC 271040768 & No & Outside\\
		Kepler-36 b,c & TIC 350810590 & No & Shallow\\
		Kepler-38 b & TIC 158316612 & No & Outside\\
		Kepler-47 b,c,d & TIC 271548206 & No & b shallow; c,d outside\\
		Kepler-51 b,c,d & TIC 27846348 & \textbf{Yes} - b,d; No - c & c shallow\\
		Kepler-63 b & TIC 299158887 & \textbf{Yes} & -\\
		Kepler-65 b,c,d,e & TIC 121731834 & No & b,c,d shallow; e outside\\
		Kepler-68 b,c & TIC 417676622 & \textbf{Yes} -b; No c & c shallow\\
		Kepler-78 b & TIC 270701667 & No & Shallow\\
		Kepler-79 b,c,d,e & TIC 239306681 & No & All shallow\\
		Kepler-83 b,c,d & TIC 123416515 & No & All shallow\\
		Kepler-89 b,c,d,e & TIC 273231214 & \textbf{Yes} - d; No - b,c,e & b,c shallow; e outside\\
		Kepler-91 b & TIC 352011875 & No & Shallow\\
		Kepler-93 b,c & TIC 137151335 & No & b shallow; c outside\\
		Kepler-96 b & TIC 169081296 & \textbf{Yes} & -\\
		Kepler-122 b,c,d,e & TIC 122714267 & No & b,c,d shallow; e outside\\
		Kepler-138 b,c,d & TIC 159376971 & No & b,c shallow; d outside\\
		Kepler-289 b,c,d & TIC 273234825 & \textbf{Yes} - d; No - b,c & b shallow; c outside\\
		Kepler-297 b,c & TIC 48304302 & No & Both shallow\\
		Kepler-381 b,c & TIC 164884235 & No & Both shallow\\
		Kepler-396 b,c & TIC 27769688 & \textbf{Yes} &  - \\
		Kepler-408 b & TIC 48450369 & No & Shallow\\
		Kepler-409 b & TIC 270619260 & No & Shallow\\
		Kepler-411 b,c,d & TIC 399954349 & \textbf{Yes} - c,d; No b & b shallow\\
		Kepler-412 b & TIC 158170594 & \textbf{Yes} & - \\
		Kepler-413 b & TIC 298969838 & No & Outside\\
		Kepler-448 b,c & TIC 169461816 & \textbf{Yes} - b; No - c & c outside\\
		Kepler-453 b & TIC 164457525 & No & No longer transiting\\
		Kepler-462 b & TIC 269263577 & No & Shallow\\
		Kepler-508 b & TIC 271671025 & No & Shallow\\
		Kepler-538 b & TIC 28227113 & \textbf{Yes} & - \\
		Kepler-1084 b & TIC 267749737 & No & Shallow\\
		Kepler-1244 b & TIC 123447592 & No & Shallow\\
		Kepler-1517 b & TIC 158555987 & \textbf{Yes} & - \\
		Kepler-1647 b & TIC 170344769 & No & Outside\\
		Kepler-1661 b & TIC 164886585 & No & Outside\\
		PH1 b & TIC 170348142 & No & Outside
	\end{tabular}
\end{table*}


\bsp	
\label{lastpage}
\end{document}